\documentclass{article}
\usepackage{spconf,amsmath,graphicx}

\usepackage{enumitem}
\setlist{nosep, leftmargin=14pt}

\usepackage{mwe} 

\title{Decoupling Shape and Density for Liver Lesion Synthesis Using Conditional Generative Adversarial Networks}
\name{Dario Augusto Borges Oliveira}
\address{IBM Research\\ Rua Tutoia 1157, Vila Mariana, Sao Paulo, Brazil}

\begin{document}
%
\maketitle
\begin{abstract}
Lesion synthesis received much attention with the rise of efficient generative models for augmenting training data, drawing lesion evolution scenarios, or aiding expert training. The quality and diversity of synthesized data are highly dependent on the annotated data used to train the models, which not rarely struggle to derive very different yet realistic samples from the training ones. That adds an inherent bias to lesion segmentation algorithms and limits synthesizing lesion evolution scenarios efficiently. This paper presents a method for decoupling shape and density for liver lesion synthesis, creating a framework that allows straight-forwardly driving the synthesis. We offer qualitative results that show the synthesis control by modifying shape and density individually, and quantitative results that demonstrate that embedding the density information in the generator model helps to increase lesion segmentation performance compared to using the shape solely.    
\end{abstract}
\begin{keywords}
Liver lesion synthesis, Generative adversarial networks, Controllable synthesis
\end{keywords}
\section{Introduction}
\label{sec:intro}

Early detection and efficient monitoring are crucial for treatment success for different kinds of cancer, and the visual inspection of CT images is one of the typical supporting procedures. Automated liver lesions identification in CT exams is especially challenging due to the high variability of shapes, densities, locations, and heterogeneity observed \cite{review1,oliveira3, oliveira4}. In this context, efficient synthesis tools might serve as a tool for creating synthetic yet realistic lesion evolution scenarios or deriving a varied set of samples for training more robust lesion segmentation and classification models.

An extensive list of approaches has been proposed for lesion synthesis, as covered in different reviews for lesion synthesis found in the literature \cite{review1,review2}. The current state-of-art for lesion synthesis often use generative adversarial networks (GANs), as reported in \cite{medgan,review3}. One of the challenges observed in lesion synthesis is how to increase the synthesis control to derive lesions with desired shapes, densities, or structures. 

This paper proposes a method for decoupling shape and density for liver lesion synthesis, allowing straight-forward synthesis control. Our results illustrate the control mechanism by modifying shape and density individually and demonstrate that the synthesized data can help increasing lesion segmentation performance on the publicly available LIST dataset in comparison to using only shape information, which indirectly points to more efficient synthesis. 


\section{Methodology}
\label{sec:method}

Our method implements shape and density decoupling for liver lesion synthesis by proposing a modification in the generator of conditional GANs, as exposed in the following sub-sections. 

\subsection{Lesion Synthesis Using Conditional GANs} 

Generative Adversarial Networks (GANs) are composed of two networks: a generator ($G$) that synthesizes images $y$, and a discriminator ($D$) that learns to classify a given image as synthetic or real. Such networks are trained in an adversarial scheme: while $G$ learns to produce realistic images to fool $D$, $D$ tries to discriminate images generated by $G$ from real images. Formally, given any data distribution $p_{data}(x)$, the generator $G$ learns a distribution $p_{model}(w)$ such that the discriminator can hardly distinguish between samples coming from $p_{data}(x)$ and $p_{model}(w)$. 

Conditional GANs (cGANs) are an extension of GANs where the input to the discriminator consists of samples from two domains ($x$ and $y$), and the generator synthesizes samples from one of those domains (say $y$). The loss function for conditional GANs is expressed by Equation~\ref{equ:cgansobjective}.
    
\begin{equation} \label{equ:cgansobjective}
\begin{split}
    \mathcal{L}_{cGAN}(G,D) = E_{x,y\sim p_{data}(x,y)}[log D(x,y)] \\
    +E_{x\sim p_{(x)},z\sim p_{(z)}}[log(1 - D(x,G(x,z))]
\end{split}
\end{equation}
    
The solution to Equation~\ref{equ:cgansobjective} is implemented by training the generator $G$ and discriminator $D$ alternately. The discriminator training uses images produced by the last trained generator and real ones. The generator is trained using the outcome of the previously trained discriminator and learns to synthesize realistic images. At the end of several training cycles, the generator is supposedly capable of producing images that the discriminator cannot distinguish from real ones. 

\subsection{Generator Modification} 

\begin{figure}
\centering
\includegraphics[width=0.49\textwidth]{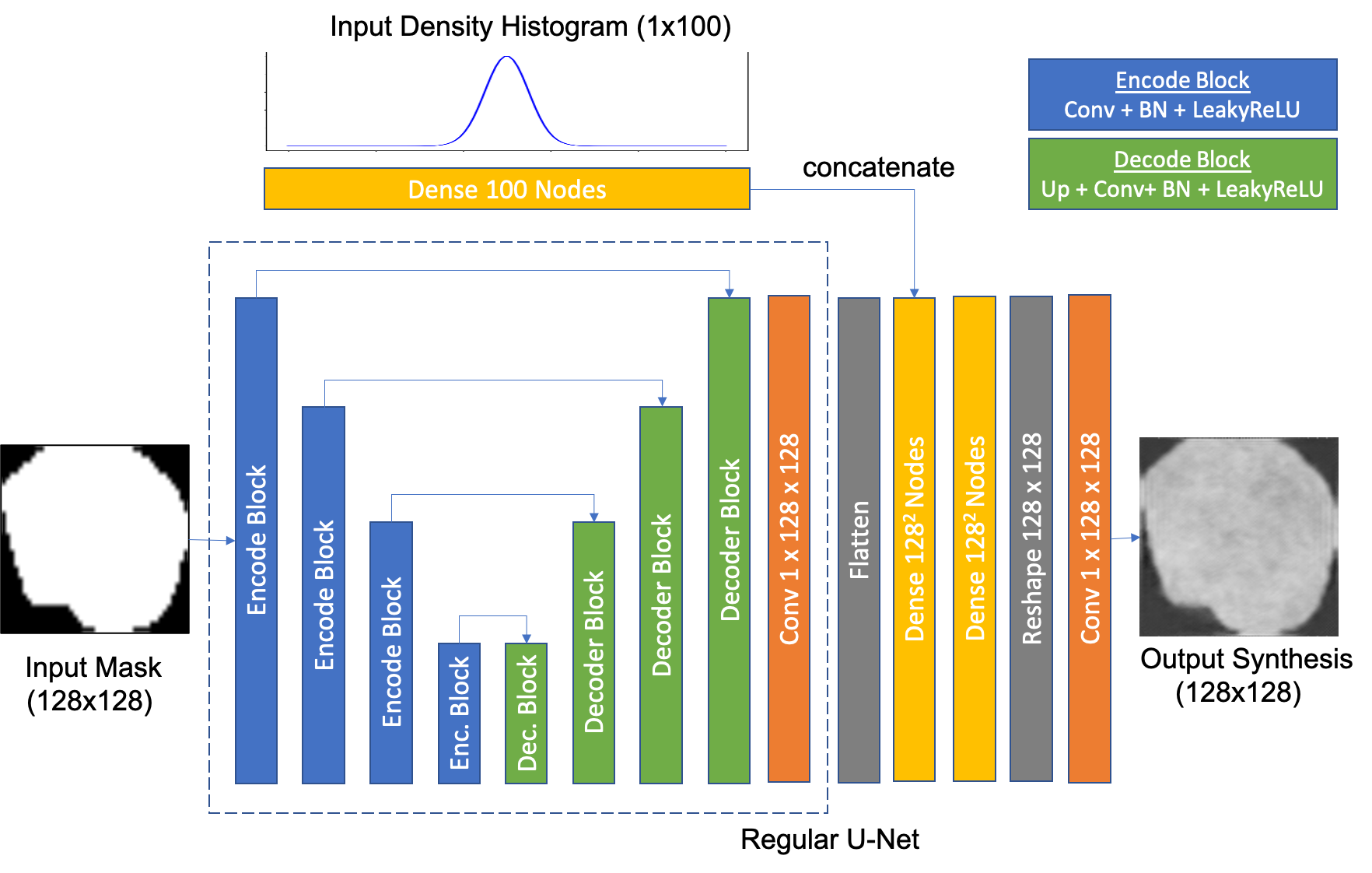}
\caption{For decoupling shape and density, we propose to create a branch for handling densities and connecting it to the original U-Net architecture using dense layers. The synthesis outcome will depend not only on the shape information but also on the densities.} \label{modified-unet}
\end{figure}

A widespread use of cGANs for lesion synthesis consists of merely using a given mask and train the generator to create characteristic lesion textures respecting the mask shape \cite{Abhishek2019Mask2LesionMA}. Here, to enable controlling the density and shape of lesion synthesis, we propose modifying the U-Net Pix2Pix generator architecture and embed a branch for handling an input histogram of densities. We expect this branch acts as an encoder or densities and structures that would enable creating different lesion appearances for the same shape and derive more diverse sample synthesis.

Figure \ref{modified-unet} depicts a regular U-Net generator architecture and our proposed modification with a branch for densities connected to the mask decoder using dense layers. One can notice the regular U-Net with encode blocks, consisting of convolutional with stride of two, batch normalization and leaky linear unit activation, followed by decoder blocks with skip connections, consisting of upsampling, convolutional layers, batch normalization and leaky linear unit activation. In our modified version, a histogram represents the densities with 100 bins, which is fed into a dense layer with 100 nodes. We flatten the last U-Net decode block output, connect to a dense layer that is concatenated to the histogram dense layer, which is finally connected to a dense layer with 128x128 nodes reshaped into a 128x128 matrix and fed to the output convolutional layer. 

In this generator, input samples consist of a data pair: an input mask that encodes the lesion shape and a histogram of densities that encodes the densities observed in the synthesized lesion. In training mode, an input segmented lesion is decomposed in a mask and a histogram of densities within the mask, and the network learns to reconstruct the original lesion out of these input data. In testing mode, modifying the shape or the histogram will enable synthesizing different lesions accordingly.

\section{Experiments and Discussion}

For running our experiments, we used the public challenge LITS dataset \cite{lits}, which sets 100 exams for training and validation, and a testing set with 30 exams used with no modification for evaluating the semantic segmentation network performances. The training set comprises a total of around 10.000 liver CT slices and 4.400 samples of lesion CT slices. 

\begin{figure}
\centering
\includegraphics[width=0.45\textwidth]{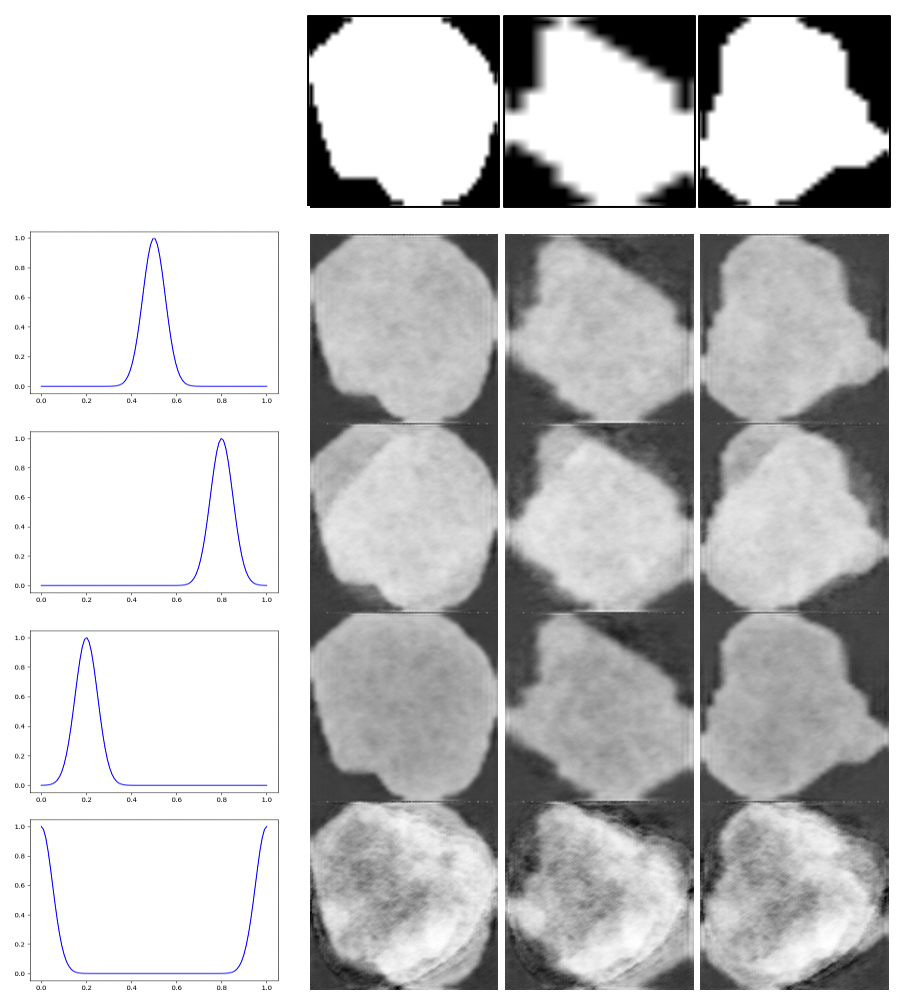}
\caption{Lesion synthesis results. The image presents three different shapes, represented by the top row, and four different patterns for densities, represented by the first column. It is possible to observe that the synthesis follows the input shapes and more interestingly, that modifying the histogram impacts the lesion appearance. } \label{synthesis-examples}
\end{figure}

We report our results from two perspectives: a visual inspection of how the manipulation of shape and density impacts the lesion synthesis and a quantitative assessment of the improvement in a segmentation network's performance when we add the density information to the synthesis compared to merely using the shape. 


\subsection{Lesion Synthesis} 

We trained two models for lesion synthesis: a baseline using a regular Pix2Pix to synthesize mask into lesions, and our modified model to synthesize masks with density histograms into lesions. For training them, we run 150 epochs using Adam optimizer with an initial learning rate of 0.0002 and a momentum of 0.5. The parameters for the generator loss function were set to GAN weight=1 and L1 weight=100. The parameters for loss functions were set as the default proposed in the paper \cite{pix2pix}.

We combined different input masks and histograms and fed into our trained model to demonstrate how the synthesis reacts to different shapes and histograms. The results are depicted in Figure \ref{synthesis-examples}. We can observe that the network respects the input masks shapes in all examples, and more interestingly, the variation observed in the histograms derive different realistic lesion patterns. They are not a mere histogram transformation, and the network learned to create textures for more homogeneous or heterogeneous lesions. In the first, second, and third rows, one can observe a simple variation in the average density, and in the fourth row, we notice that creating more modals in the histogram enables creating more complex lesion structures in the synthesized sample. 

This framing allows controlling lesion synthesis in two ways: modifying the synthetic lesion shape by adjusting the input mask and modifying the lesion appearance by changing the input histogram of densities.   

\subsection{Lesion Segmentation}

\begin{figure*}
\centering
\includegraphics[width=0.7\textwidth]{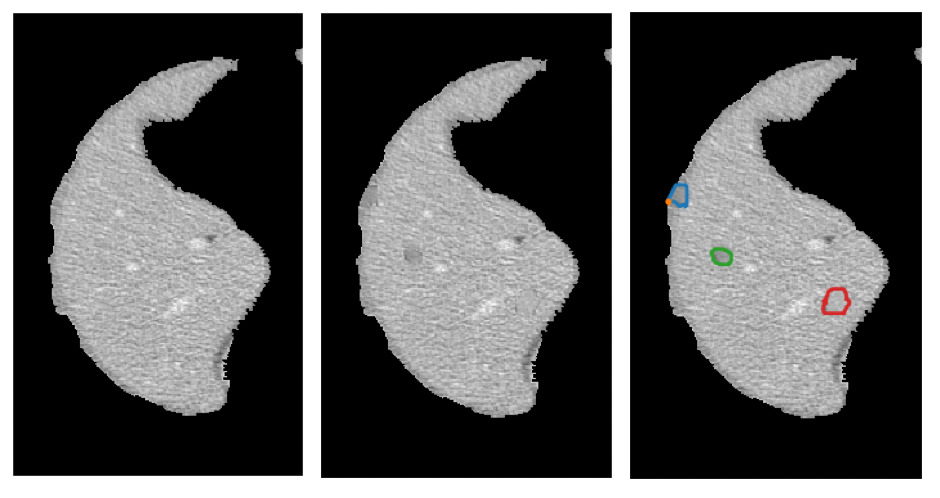}
\caption{Lesion implant procedure in healthy liver CT slices. Each synthesized lesion is rotated and scaled at random and blended to the liver parenchyma.} \label{implant}
\end{figure*}

To better understand the implications and possible impacts of this framing, we trained models using the popular PSP-Net \cite{pspnet} multi-scale semantic segmentation architecture to evaluate different training set configurations. First, we used the original LITS dataset to set a ceiling performance value. Then, we derived two synthetic datasets, one using the regular U-Net mask to lesion schema, and the other using our approach. The datasets are created by implanting lesions in existing healthy liver CT slices, up to the same amount available in the original training data, as in \cite{oliveira}. That allows us to compare the performance of models trained using synthesized data from regular Pix2Pix, our proposed approach, and the ceiling performance achieved using real data. We selected lesions shapes and histogram of densities at random. We also rotated and scaled the lesion at random and picked a random coordinate for implanting the lesion. 

The impact of the more diverse training set created using our approach into the segmentation network performance was encouraging and points to the benefit of controlling the lesion appearance using the histogram of densities. We observed an increase in performance of around \textbf{6\%} in F1-score compared to using only the mask (40.13\% versus 34.09\%) with a regular U-Net generator setup, and a training set using only synthetic data derived 67\% of the performance observed when using real data. This result points to a more diverse lesion dataset using density information for controlling the synthesis.

A critical remark is that our focus in this paper is to study the impact of decoupling shape and density for lesion synthesis control. With that in mind, we envisage that exploring different models for synthesis would potentially derive better quality lesion synthesis outcome, and we also understand that other models for liver lesion segmentation deliver better results than the ones observed in our baseline. Still, we consider the models used adequate for evaluating our core contribution. 

\begin{table}
\centering
\caption{Liver lesion segmentation F1 score for different training sets, using only synthetic data for synthesis from mask, synthesis from mask+density, and using only original data.}\label{tab1}
\begin{tabular}{c|c}
 &  { PSP F1-Score}    \\
\hline
\hline
{Original Data}  & 0.5996  \\
\hline
{Pix2Pix Synthesis from Mask}  & 0.3409  \\
\hline
{Pix2Pix Synthesis from Mask+Density}  & \textbf{0.4013} \\
\hline
\end{tabular}
\label{table-results}
\end{table}

\section{Conclusions}

This paper presents a method for improving lesion synthesis control by decoupling lesion shape and density using conditional adversarial networks. We proposed modifying the popular Pix2Pix architecture generator and created a branch for handling densities in addition to the branch for the widespread mask-to-lesion synthesis. 

We presented two different sets of results for evaluating our model: one for the visual inspection of synthesis control and the other for the numerical impact in performance considering a commonly used semantic segmentation network. First, we showed that our model could create synthetic lesions consistent with the input shape and densities, represented by a lesion mask and a histogram of densities, respectively. Variations on the densities enable more or less dense lesion synthesis and complex and heterogeneous lesion synthesis when different modals are observed in the histogram. 

Then, we trained semantic segmentation networks using different training sets: one with original data, which should deliver the ceiling performance, and two using only synthetic data considering the original Pix2Pix setup for synthesis from masks, and our design that also considers the densities. We reported a gain of 6\% in performance in our model, which points to a more diverse training set.

As further research, we intend to explore more robust models for data synthesis and synthesize 3D samples to verify if the findings will be consistent with those observed in this study.






\bibliographystyle{IEEEbib}
\bibliography{strings,refs}

\end{document}